\documentclass[a4paper,aps,twocolumn,prb,showpacs,amsmath,amssymb,superscriptaddress,reprint]{revtex4-2}

\usepackage{graphicx}
\usepackage{dcolumn}% 
\usepackage{bm}%
\usepackage[colorlinks = true,
            linkcolor = blue,
            urlcolor  = blue,
            citecolor = blue,
            anchorcolor = blue]{hyperref}
\usepackage[english]{babel}
\usepackage{natbib}
\usepackage{color}
\usepackage{amsmath}
\usepackage{amssymb}
\usepackage{subfigure}

\begin{document}
\renewcommand{\figurename}{FIG.}

\title{Hard x-ray angle-resolved photoemission from a buried high-mobility electron system}

\author{M.~Zapf}
\affiliation{Physikalisches Institut and Würzburg-Dresden Cluster of Excellence ct.qmat, Julius-Maximilians-Universität, 97074 Würzburg, Germany}
\author{M.~Schmitt}
\email{mschmitt@physik.uni-wuerzburg.de}
\affiliation{Physikalisches Institut and Würzburg-Dresden Cluster of Excellence ct.qmat, Julius-Maximilians-Universität, 97074 Würzburg, Germany}
\author{J.~Gabel}
\affiliation{Physikalisches Institut and Würzburg-Dresden Cluster of Excellence ct.qmat, Julius-Maximilians-Universität, 97074 Würzburg, Germany}
\affiliation{Diamond Light Source Ltd., Didcot, Oxfordshire OX11 0DE, United Kingdom}
\author{P.~Scheiderer}
\affiliation{Physikalisches Institut and Würzburg-Dresden Cluster of Excellence ct.qmat, Julius-Maximilians-Universität, 97074 Würzburg, Germany}
\author{M.~St\"ubinger}
\affiliation{Physikalisches Institut and Würzburg-Dresden Cluster of Excellence ct.qmat, Julius-Maximilians-Universität, 97074 Würzburg, Germany}
\author{B.~Leikert}
\affiliation{Physikalisches Institut and Würzburg-Dresden Cluster of Excellence ct.qmat, Julius-Maximilians-Universität, 97074 Würzburg, Germany}
\author{G.~Sangiovanni}
\affiliation{Institut für Theoretische Physik und Astrophysik and Würzburg-Dresden Cluster of Excellence ct.qmat, Julius-Maximilians-Universität, 97074 Würzburg, Germany}
\author{L.~Dudy}
\affiliation{Physikalisches Institut and Würzburg-Dresden Cluster of Excellence ct.qmat, Julius-Maximilians-Universität, 97074 Würzburg, Germany}
\affiliation{SOLEIL Synchrotron, 91190 Saint-Aubin, France}
\author{S.~Chernov}
\affiliation{Insitut für Physik, Johannes Gutenberg-Universit\"at, 55122 Mainz, Germany}
\author{S.~Babenkov}
\affiliation{Insitut für Physik, Johannes Gutenberg-Universit\"at, 55122 Mainz, Germany}
\author{D.~Vasilyev}
\affiliation{Insitut für Physik, Johannes Gutenberg-Universit\"at, 55122 Mainz, Germany}
\author{O.~Fedchenko}
\affiliation{Insitut für Physik, Johannes Gutenberg-Universit\"at, 55122 Mainz, Germany}
\author{K.~Medjanik}
\affiliation{Insitut für Physik, Johannes Gutenberg-Universit\"at, 55122 Mainz, Germany}
\author{Yu.~Matveyev}
\affiliation{DESY Photon Science, 22607 Hamburg, Germany}
\author{A.~Gloskowski}
\affiliation{DESY Photon Science, 22607 Hamburg, Germany}
\author{C.~Schlueter}
\affiliation{DESY Photon Science, 22607 Hamburg, Germany}
\affiliation{Diamond Light Source Ltd., Didcot, Oxfordshire OX11 0DE, United Kingdom}
\author{T.--L.~Lee}
\affiliation{Diamond Light Source Ltd., Didcot, Oxfordshire OX11 0DE, United Kingdom}
\author{H.--J. Elmers}
\affiliation{Insitut für Physik, Johannes Gutenberg-Universit\"at, 55122 Mainz, Germany}
\author{G. Sch\"onhense}
\affiliation{Insitut für Physik, Johannes Gutenberg-Universit\"at, 55122 Mainz, Germany}
\author{M.~Sing}
\affiliation{Physikalisches Institut and Würzburg-Dresden Cluster of Excellence ct.qmat, Julius-Maximilians-Universität, 97074 Würzburg, Germany}
\author{R.~Claessen}
\affiliation{Physikalisches Institut and Würzburg-Dresden Cluster of Excellence ct.qmat, Julius-Maximilians-Universität, 97074 Würzburg, Germany}

\date{\today}

\begin{abstract}
Novel two-dimensional electron systems at the interfaces and surfaces of transition-metal oxides recently have attracted much attention as they display tunable, intriguing properties that can be exploited in future electronic devices. Here we show that a high-mobility quasi-two-dimensional electron system with strong spin-orbit coupling can be induced at the surface of a KTaO$_3$ (001) crystal by pulsed laser deposition of a disordered LaAlO$_3$ film. The momentum-resolved electronic structure of the buried electron system is mapped out by hard x-ray angle-resolved photoelectron spectroscopy. From a comparison to calculations, it is found that the band structure deviates from that of electron-doped bulk KTaO$_3$ due to the confinement to the interface. Fermi surface mapping shows a three-dimensional, periodic intensity pattern consistent with electron pockets of quantum well states centered around the $\Gamma$ points and the expectations from a Fourier analysis-based description of photoemission on confined electron systems. From the $k$ broadening of the Fermi surface and core-level depth profiling, we estimate the extension of the electron system to be at least 1\,nm but not much larger than 2\,nm, respectively.      
\end{abstract}

\maketitle
\section{Introduction}
Ternary metal oxides of the perovskite type with the structural formula ABO$_3$ exhibit a great manifold of interesting ground states ranging from multiferroic insulators and magnetic semiconductors to superconducting metals.
One of the most prominent phenomena studied in these compounds in recent years is the formation of metallic quasi-two-dimensional electron systems (q2DESs) at the surfaces and heterointerfaces of band insulating members of this family of materials~\cite{mannhart2010}.  

Most of the q2DESs studied so far have been based on the prototypical perovskite SrTiO$_3$~(STO) and have mostly been attributed to arise from $n$-doping spatially confined to a few crystal layers. Many routes for the generation of such metallic electron systems in STO have now been identified. Examples are negative charge accumulation induced by the electric field-effect, electron transfer driven by a polar discontinuity at a heterointerface, and local doping by impurity atoms or changes of the oxygen stoichiometry from ABO$_3$ to ABO$_{3-\delta}$ that is caused either by chemical reduction or irradiation with intense photon beams~\cite{ohtomo2004,ueno2008,jalan2010,chen2011,meevasana2011,lee2012,gabel2017}. This leads to a diversity of intriguing features and controls in such systems, including high charge carrier densities and mobilities, the unusual coexistence of  ferromagnetism and superconductivity, charge writing, tunability of the Rashba interaction, and gate-controlled conductivity that have triggered broad activities in both fundamental and applied research~\cite{ohtomo2004,thiel2006,reyren2007,xie2010,caviglia2010,li2011,bert2011,chen2013}.

Obviously, with the large variety of tantalizing phenomena found in STO q2DESs, there is a quest to introduce similar electron systems into other, more exotic materials that may exhibit superior electronic transport properties and have the potential to evoke new functionalities in the q2DESs through the inherent properties of their parent compounds. One promising parameter that can be exploited is the large spin-orbit coupling~(SOC) of high-$Z$ elements (see also Appendix~\ref{app:SOCdim}), which, for instance, is very interesting for spintronics applications. Integrated into a suitable microelectronic structure, a q2DES with strong SOC offers control over the spin state of the charge carriers in a two-dimensional conduction channel by applying electric fields~\cite{zutic2004}.

A promising host material for the stabilization of such an electron system with considerably large SOC is the 5$d$-transition metal oxide KTaO$_3$~(KTO). 
Experiments under ultrahigh vacuum conditions on cleaved single crystals verified the formation of a metallic q2DES in the uppermost lattice layers of the (001) and (111) surfaces of KTO~\cite{king2012,santander-syro2012,bareille2014,bruno2019}. Recently, intriguing properties of this q2DES like superconductivity and the Edelstein effect have been demonstrated \cite{vicente-arche2021,liu2021,ojha2021}. Most likely induced through $n$ doping resulting from the formation of oxygen vacancies~(V$_\text{O}$)~\cite{king2012} at the surfaces, these q2DESs quickly decay when the crystals are exposed to air. However, for a comprehensive experimental characterization and also for use in prospective technological applications it is necessary for the q2DESs to be stable under ambient conditions.

Here we demonstrate the preparation of such q2DESs that are stable at room temperature and in ambient atmosphere thanks to suitable nm-thin protecting layers. This allows us to study their macroscopic and microscopic electronic properties in identical geometry, i.e., with capping layers grown under identical conditions, by electrical transport measurements and hard x-ray photoemission spectroscopy, respectively. We find significantly higher electron mobilities than previous studies of KTO-based electron systems. From core-level depth profiling and momentum-resolved mapping using hard x-rays, we infer that the electron system is only about 2\,nm thick and, in fact, quasi-two-dimensional in nature.

\section{Experimental Details}
The heterostructure samples were prepared by pulsed laser deposition~(PLD) from a LaAlO$_3$ (LAO) crystal at an oxygen partial pressure of 10$^{-7}$--10$^{-6}$\,mbar onto a KTO~(001) substrate ($a_\text{KTO}=3.989$\,{\AA}~\cite{zhurova2000}) using laser fluencies of 1.1--1.8\,Jcm$^{-2}$. Epitaxial growth is inhibited by the large lattice mismatch of 5.2\,\% \cite{zhang2017} and would also lead to a reversed polarity with respect to the substrate, which might affect the electronic properties of the 2DES. Therefore, disordered LAO~(dLAO) films were deposited while the substrates were kept at room temperature. Electronic transport and Hall effect measurements were conducted in a Quantum Design physical property measurement system. Hard x-ray photoemission measurements in angle-integrated~(HAXPES) and angle-resolved~(HARPES) modes with photon energies in the range of 2500--3500\,eV were performed at temperatures of $T \approx 55$\,K at beamline I09 at Diamond Light Source with a VG Scienta EW4000 hemispherical electron analyzer featuring a wide acceptance angle of about $\pm 30^\circ$ along the analyser slit, resulting in a $k_{\parallel}$ resolution of about 0.16\,{\AA}$^{-1}$ and an energy resolution of about 350\,meV.

Further HARPES measurements were conducted (see Appendix~\ref{app:ToF-mic}) at beamline P22 of PETRA III~\cite{schlueter2019} at temperatures of $T\approx30$\,K with a momentum-space time-of-flight (ToF) microscope with a front lens optimized for low aberrations at very high excitation energies~\cite{medjanik2019}. The energy and $k$ resolutions with this instrument amount to about 600\,meV (at 5.2\,keV photon energy) and 0.025\,{\AA}$^{-1}$, respectively.

\section{Preparation of a buried quasi-two-dimensional electron system and electronic transport}
\begin{figure*}
 \includegraphics[width=\textwidth]{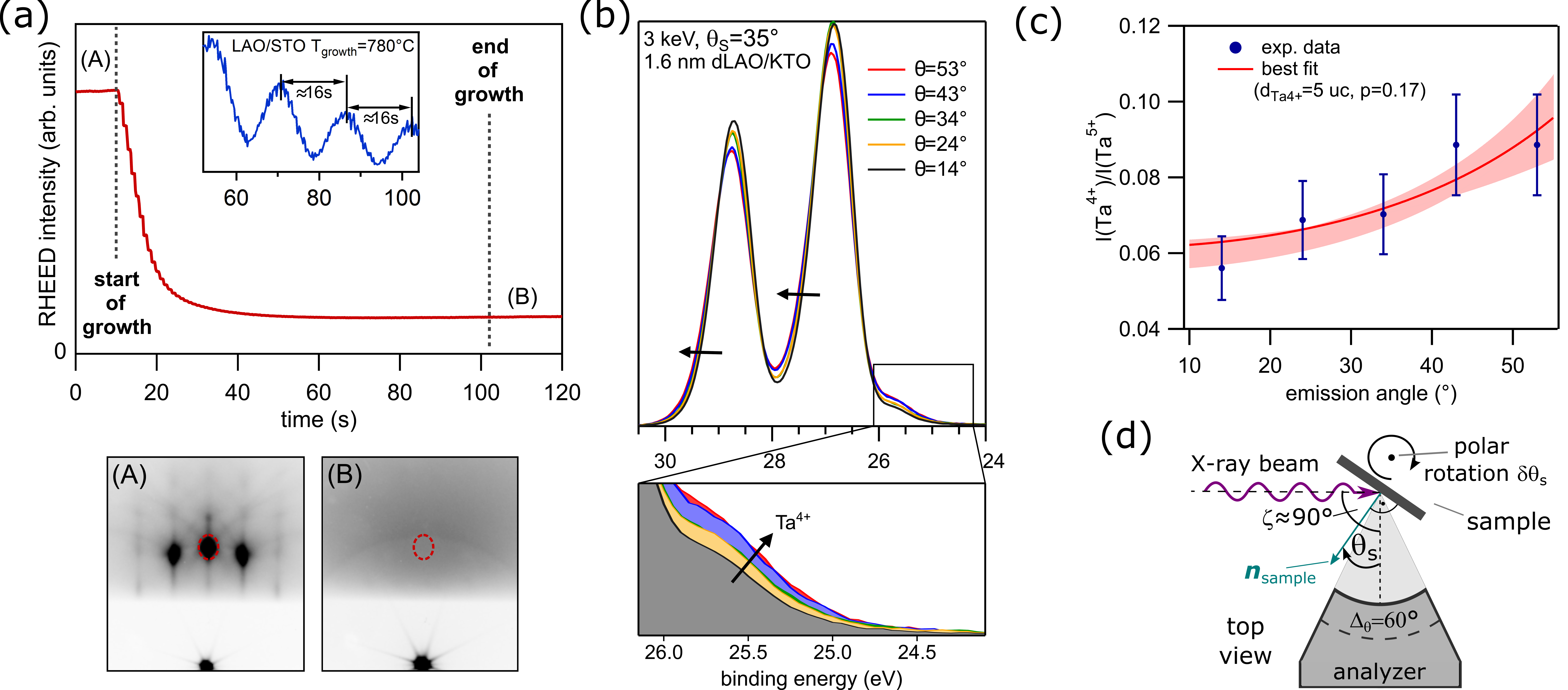}
 \caption{\label{fig:growth_depth-profile}~(a)~Upper panel: RHEED intensity of the specular spot (dashed red circle) during deposition of dLAO at room temperature (inset: thickness calibration by comparison with epitaxial LAO growth on STO). Lower panel: RHEED patterns before (A) and after growth (B). (b)~Background-corrected Ta~4$f$ spectra (top) and close-up of the Ta$^{4+}$ emission region (bottom). A distinct angle-dependence of the Ta$^{4+}$ spectral weight and an asymmetry of the Ta~4$f$ lines towards higher binding energies are observed. (c)~A detailed analysis based on a model fit of the Ta$^{4+}$/Ta$^{5+}$ intensity ratio as described in Ref.~\cite{sing2009} yields a q2DES extension of $\approx 20$\,{\AA}. (d)~Scheme of the experimental geometry of the HAXPES setup at endstation I09 at the Diamond Light Source.}
\end{figure*}

The q2DES at the surface of KTO has been suggested to arise from oxygen vacancies that act as electron donors~\cite{king2012}. Here we create such a V$_\text{O}$-based metallic electron system in a KTO substrate by the deposition of an LAO film by PLD in oxygen partial pressures below $p_{O_2} = 10^{-6}$\,mbar and with the substrate held at room temperature. Ta$^{5+}$ ions with an empty $d$ shell are expected to be reduced to Ta$^{4+}$ with a 5$d^1$ configuration, and metallicity of the KTO surface emerges similar to the functionalized STO surfaces~\cite{chen2011}.

The monitoring by reflection high-energy electron diffraction~(RHEED) of the LAO deposition, exemplarily shown in Fig.~\ref{fig:growth_depth-profile}(a), signifies an exponential damping of the coherent intensity of the specular diffraction spot [integrated over the dashed red circle in diffraction pattern~(A)] with deposition time. Accordingly, the sharp RHEED pattern of KTO before growth [image~(A)] vanishes entirely and only a homogeneous intensity distribution due to incoherent scattering remains after growth of the LAO film [image~(B)], signaling its non-crystalline, disordered nature. The thickness of the deposited oxygen deficient dLAO film, $t_{\text{dLAO}}$, is determined by the number of laser pulses in the growth process. The corresponding thickness-pulse number relationship was calibrated, as exemplarily shown in the inset of Fig.~\ref{fig:growth_depth-profile}(a), by counting the number of pulses necessary to grow one unit cell, corresponding to one RHEED oscillation, of an epitaxial, crystalline LAO film at the same oxygen background pressure and laser fluence on an STO substrate. According to this calibration the thickness of the dLAO film of the sample is estimated to be $t_{\text{dLAO}} = 2.4$\,nm.

To evidence the formation of the q2DES and its extension, we employed angle-dependent HAXPES that allows us to tune the probing depth and thus obtain a vertical profile of the electronic and chemical properties of a sample. The method is based on the fact that the inelastic mean-free path $\lambda_e$ of the photoelectrons in a solid effectively varies as $\lambda_e \cos \theta $, where $\theta$ is the emission angle of the photoelectrons with respect to the surface normal. Because the O~2$s$ core level contributes some background to the Ta~4$f$ line shape (not shown), in particular, in the Ta$^{4+}$ region, the background was removed by fitting the broad O~2$s$ lineshape with a Gaussian and subtracting it along with a Shirley background \cite{shirley1972}. The corrected angle-dependent Ta~4$f$ spectra are displayed in Fig.~\ref{fig:growth_depth-profile}(b). As can be judged already from the small but nevertheless clear angle dependence of the Ta$^{4+}$ weight that increases with the surface sensitivity of the measurements, the majority of the charge carriers is located near the interface with a confinement length somewhat smaller than the inelastic mean free path. A detailed analysis based on a model fit to the Ta$^{4+}$/Ta$^{5+}$ intensity ratio~\cite{sing2009}, the results of which are displayed in Fig.~\ref{fig:growth_depth-profile}(c), yields a q2DES extension of five~unit cells corresponding to $\approx 20$\,{\AA} and a fraction of 0.17 of Ta$^{4+}$ sites in the q2DES region.

In addition, an asymmetry of the core-level lines towards higher binding energies is evident in the Ta~4$f$ spectra of Fig.~\ref{fig:growth_depth-profile}(b), which becomes stronger with increasing $\theta$, i.e., higher surface sensitivity. Such an asymmetry is well-known from other surface and interface q2DESs and reflects band bending due to charge accumulation at the interface. In this space-charge region, the electrical potential induced by the effectively positively charged oxygen vacancies pulls the KTO conduction bands below the Fermi energy which thus accommodate the electrons released from the oxygen vacancies. Accordingly, the core levels in the uppermost KTO layers shift to higher binding energies, leading to the observed asymmetry of the lineshapes in the spectra.\\

\begin{figure*}
 \includegraphics[width=0.7\textwidth]{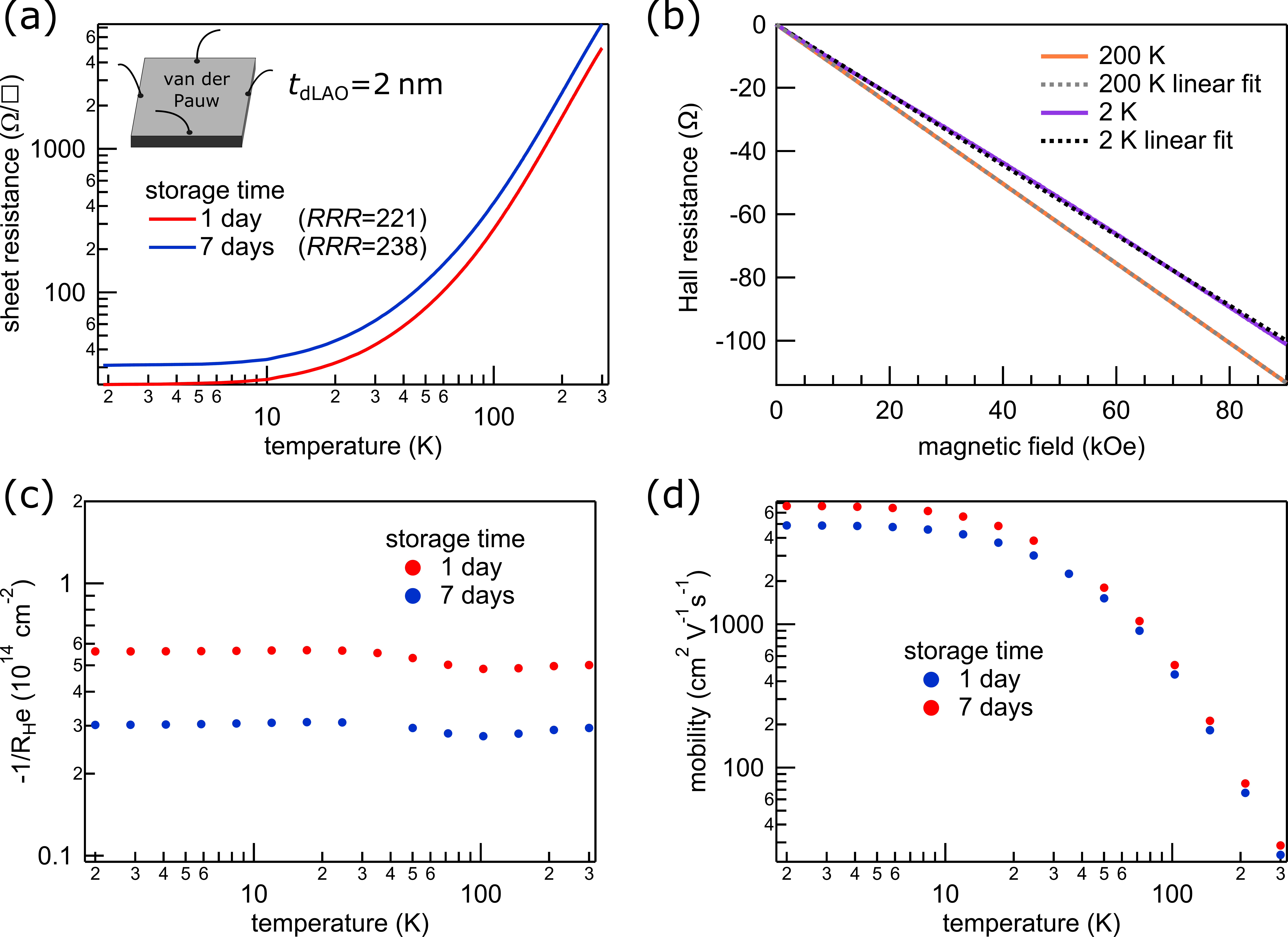}
 \caption{\label{fig:transport} (a)~Sheet resistances one~day and seven~days after deposition (stored in a desiccator at a pressure of 300\,mbar). (b)~Symmetrized Hall resistances at 2\,K and 200\,K measured one~day after deposition. (c)~Sheet carrier densities and (d)~mobilities for a sample with a film thickness $t_{\text{dLAO}}=2$\,nm measured one~day and seven~days after deposition.}
\end{figure*}

Furthermore, the successful preparation of the metallic electron system is also evidenced by the exemplary transport measurements of a 2\,nm thick dLAO/KTO heterostructure in Figs.~\ref{fig:transport}(a) and \ref{fig:transport}(b). The 4-point probe resistance measurements in Fig.~\ref{fig:transport}(a) were conducted \textit{ex situ} after contacting the samples in Van der Pauw geometry by ultrasonic wire bonding. While a bare KTO substrate is highly insulating, the dLAO/KTO heterostructure shows clear metallic behavior with the resistance decreasing with decreasing temperature (red curve). Hence, a metallic electron system is indeed induced, and is sufficiently stable under ambient conditions even after one week of storage (blue curve). The sheet resistance $R_s$ ranges from values of $10^1$ to $10^2$\,$\Omega$/$\square$ at 2\,K, being in the same range as resistance values observed in STO-based q2DESs~\cite{chen2013, chen2011, trier2013, brinkman2007, christensen2018}, with a residual resistance ratio of $R_s$(300\,K)/$R_s$(2\,K)~$>$~200.

Hall measurements up to 9\,T shown in Fig.~\ref{fig:transport}(b) evidence $n$-type, i.e., electron charge carrier transport. The linear fits as indicated by the dotted lines well describe the magnetic field dependence for the whole temperature range, pointing toward a single carrier type dominating the transport properties of the sample. Charge carrier densities of up to 6$\times10^{13}$\,cm$^{-2}$ are obtained [cf. Fig.~\ref{fig:transport}(c)]. It is noted, however, that at temperatures below 30\,K, the Hall resistance becomes slightly non-linear. At higher carrier densities, which are achieved by depositing a thicker layer of dLAO, a second conduction channel becomes occupied, resulting in a non-linear Hall resistance (not shown).

Seven days after deposition, the number of charge carriers in the system has decreased. The most plausible degradation pathway is the refilling of V$_\text{O}$ by ambient O$_2$ molecules, which dissociate at the sample surface and diffuse through the dLAO layer into KTO. The significance of this process depends on the thickness of the dLAO capping layer. A similar reduction of the charge carrier density with storage time is also known from the dLAO/STO heterostructures~\cite{trier2013, christensen2013}.

Figure~\ref{fig:transport}(d) shows an increase of the Hall mobility in dLAO/KTO with decreasing temperature, resembling the overall trend observed for doped KTO bulk crystals~\cite{wemple1965} and STO-based electron systems~\cite{tufte1967, verma2014}. Remarkably, the low-temperature mobility $\mu_0$ reaches values that are about four times larger than those reported for other dLAO/KTO studies~\cite{zhang2017} and, furthermore, one order of magnitude larger than those reported for dLAO/STO \cite{chen2011, trier2013}. With $\mu_0$ of up to 7000\,cm$^2$V$^{-1}$s$^{-1}$ at 2\,K, dLAO/KTO exhibits low-temperature mobilities comparable to the best values reported for LAO/STO heterostructures with a crystalline LAO film~\cite{xie2013, fete2015}. Furthermore, at room temperature, a $\mu_0$ of about 30\,cm$^2$V$^{-1}$s$^{-1}$ is considerably larger than those achieved in other KTO- and also STO-based q2DES systems, which only reach values around 10\,cm$^2$V$^{-1}$s$^{-1}$~\cite{christensen2018,mikheev2015,zhang2017}. We ascribe the relatively high Hall mobilities of dLAO/KTO as compared to dLAO/STO mainly to the effective masses of the electrons which are lower in the 5$d$-derived conducting states of KTO than those of the electrons in the 3$d$ states of STO~\cite{himmetoglu2016}. In addition, the depth extensions of the conduction channels and the V$_\text{O}$ distributions may differ such that the electrons in dLAO/STO are effectively more often scattered off oxygen vacancies than in dLAO/KTO.

In agreement with observations in STO-based surface and interface q2DESs, we also observe that the extra electrons are accumulated at the heterointerface, as shown by core-level photoemission in Appendix~\ref{sec:corelevel}.

\section{Hard X-ray Angle-Resolved Photoemission}
\label{sec:HARPES}
Having demonstrated the existence of a highly mobile electron system at the interface in the dLAO/KTO heterostructure and having characterized its macroscopic transport properties, we now discuss its microscopic electronic structure and compare it to that of the surface electron system on bare KTO crystals as previously investigated by very surface-sensitive, low-energy ARPES measurements performed with vacuum ultraviolet~(VUV) lights~\cite{king2012,santander-syro2012,bareille2014}. As VUV-excited photoelectrons are not able to escape through the dLAO cap, we employ higher photon energies around 3\,keV to increase the information depth of the measurements on a sample with $t_\text{dLAO}=1.6$\,nm.

\begin{figure}
 \includegraphics[width=0.38\textwidth]{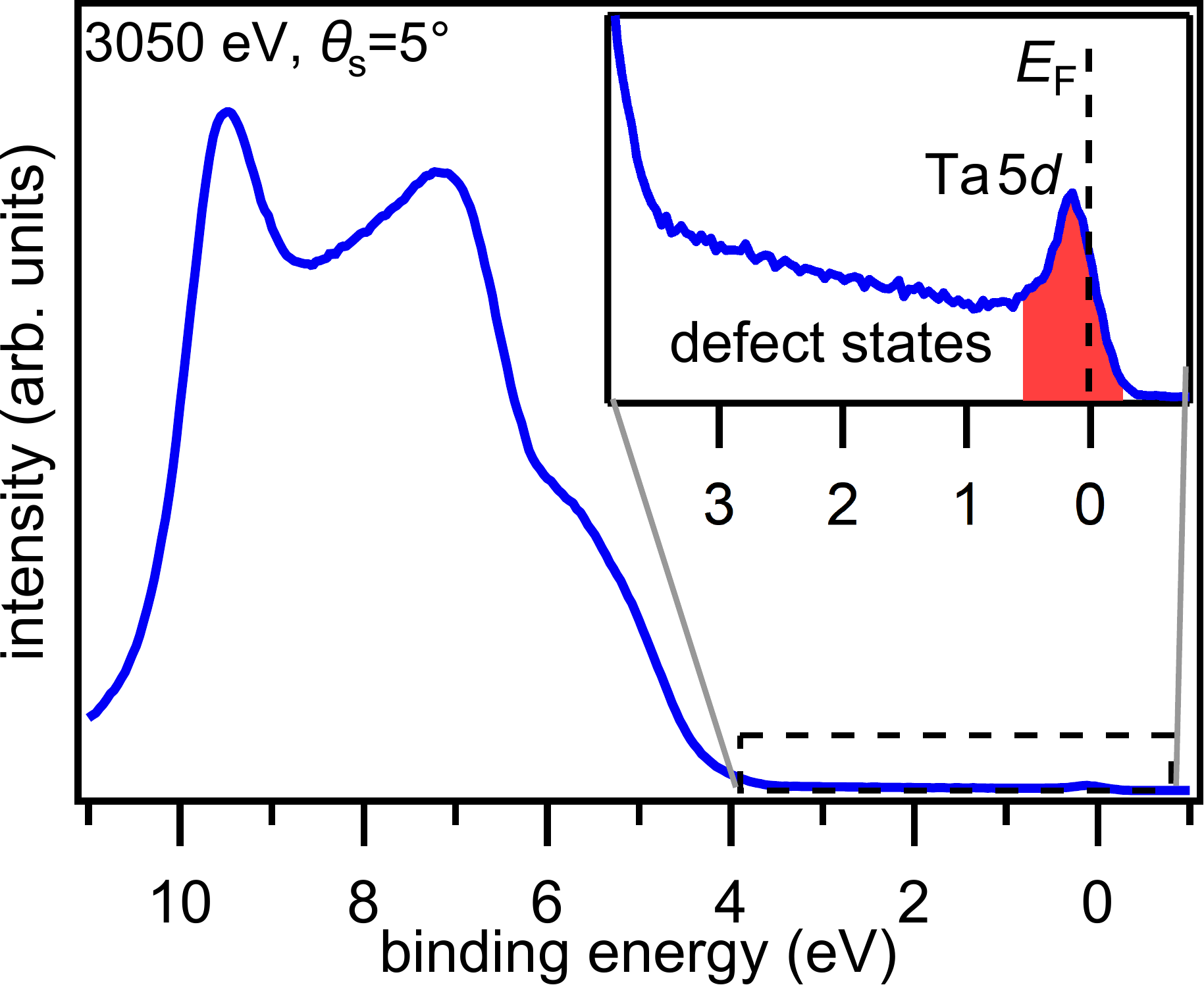}
 \caption{\label{fig:HAXPES_VB} Angle-integrated valence-band spectrum of dLAO/KTO measured with $h\nu=3050$\,eV. The mobile electrons of the q2DES residing in Ta~5$d$ states are detectable at the Fermi energy~$E_{F}$. In the KTO band gap, a broad intensity distribution is seen in the range 0.5--4\,eV, presumably associated with defect states in dLAO and KTO.}
\end{figure}

At excitation energies of several keV, the photoionization cross sections of the Ta~5$d$ states, compared to those of the Ti~3$d$ states in STO, are two orders of magnitude higher and therefore allow the Ta~5$d$ charge carriers in dLAO/KTO to be probed directly by HAXPES, although only a small fraction of Ta atoms at the interface are in the Ta$^{4+}$ oxidation state and contribute to the formation of the conducting electron system.

Figure~\ref{fig:HAXPES_VB} shows the valence band of the dLAO/KTO heterostructure recorded at a photon energy of $h\nu=3050$\,eV and integrated over the entire angular detection window of the analyzer lens, centered around an angle $\theta_s\approx5^{\circ}$ off normal emission. The part above $\approx 4$\,eV binding energy is ascribed to the valence-band states of KTO and the dLAO film. Much weaker, but well discernible in the close-up, is a distinct peak at the Fermi energy. This spectral feature is associated with the mobile Ta~5$d$ electrons that constitute the metallic electron system observed in the transport experiments.

Contiguous to this peak, in the pristine band gap of KTO, a virtually structureless distribution of spectral weight is visible between 0.5 and 4.0\,eV binding energy. We attribute this to trapped electrons in various localized defect states in both the dLAO film and the oxygen-deficient substrate.

\begin{figure*}
 \includegraphics[width=0.92\textwidth]{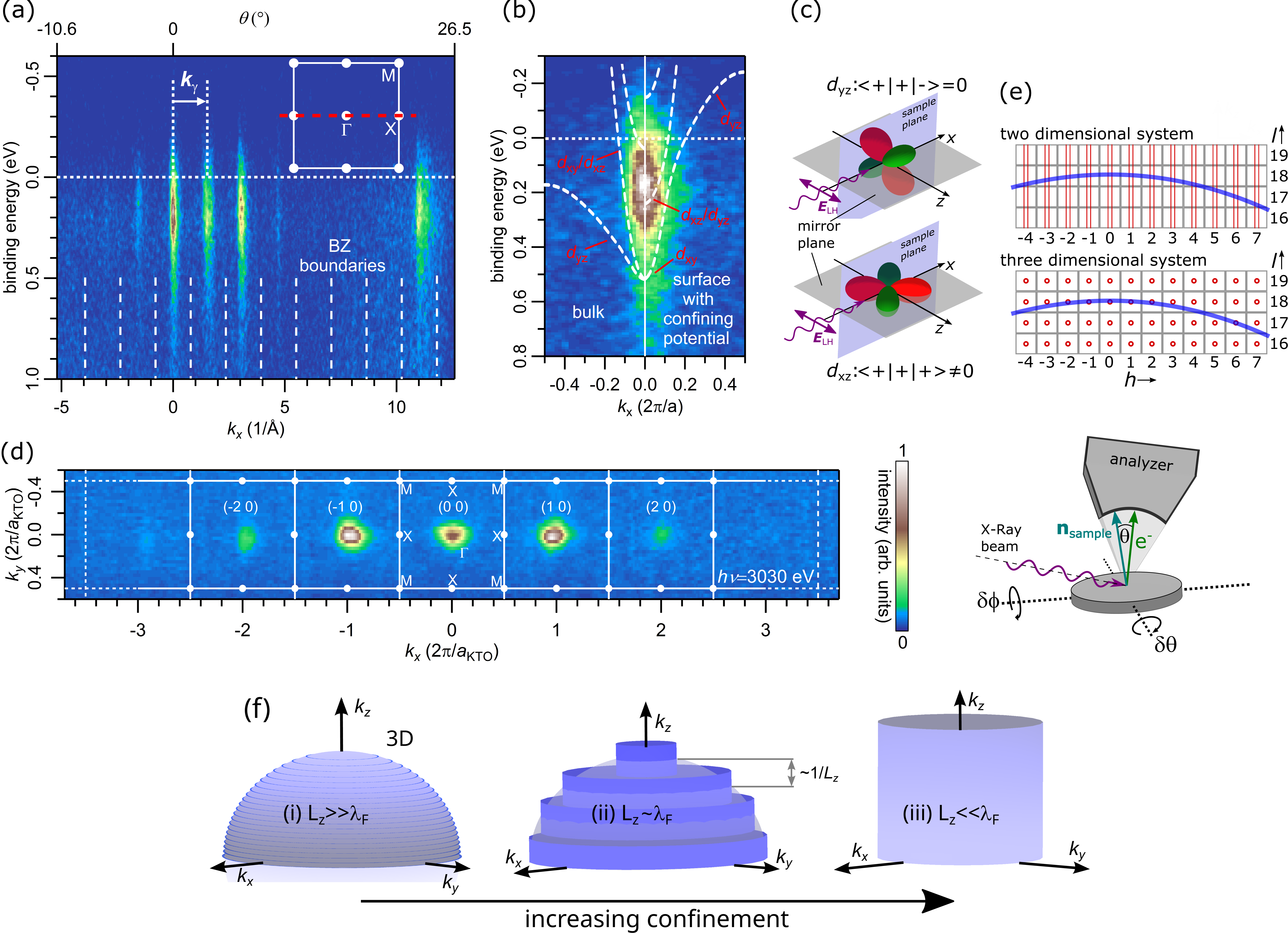}
 \caption{\label{fig:bandmap_kxky}(a)~Band map of a dLAO/KTO sample with $t_\text{dLAO}=1.6$\,nm near the chemical potential, measured along the $\Gamma-X$-direction of KTO with $h\nu=3030$\,eV. The data is not corrected for the photon momentum $k_{\gamma}$ and therefore is \textit{not} symmetric in $k_x$ direction with regard to $k_x=0$\,\AA$^{-1}$. The dashed lines in the lower part indicate the BZ boundaries in the repeated zone scheme of KTO. The BZ is shown as inset (not true to scale). (b)~Band map of BZ~$(-1~0~18)$ along $\Gamma-X$. The white lines are tight-binding calculations from Ref.~\cite{santander-syro2012} for bulk KTO (left) and for an electron system at the surface of KTO, which is confined by a wedge potential along the $z$ direction (right). (c)~Schematic orientation of the $d_{yz}$ and $d_{xz}$ orbitals with respect to sample and mirror planes for the HARPES measurements shown in (a) and (b). Symmetry considerations yield a vanishing (non-vanishing) photoemission matrix element for $d_{yz}$ ($d_{xz}$). (d)~Left: Photoemission intensity map of a 1.6\,nm dLAO/KTO sample in the $k_x-k_y$ plane, generated by integrating the spectral intensity over the binding energy range from $-0.3$\,eV to 0.5\,eV. The BZs are labeled according to their Miller indices $h$ and $k$. The measurement has been corrected for the photon momentum $k_{\gamma}$. Right: Schematic drawing of the experimental geometry for the mapping of the electron momentum in three dimensions. (e)~Cut through the repeated zone scheme of bulk KTO with the expected distribution of spectral weight for the cases of two- and three-dimensional free electron systems. (f)~Development of the Fermi surface of a free electron system with increasing confinement as defined by the relative size of its extension along the $z$ axis and the Fermi wavelength. The height of the stacked cylinders roughly reflects the width of the momentum distribution function.}
\end{figure*}

Beyond recording the angle-integrated valence-band spectrum, we also employed hard x-ray angle-resolved photoelectron spectroscopy (HARPES), which has been successfully applied to a few prototypical sample materials like W and Ga$_{1-x}$Mn$_{x}$As~\cite{gray2011,gray2012}, Ga$_{1-x}$Mn$_{x}$P~\cite{keqi2018}, Mo and TiTe$_2$~\cite{babenkov2019}, and LaB$_6$~\cite{rattanachata2021}.
A prerequisite for the application of HARPES is that the number of direct transitions excited in the photoemission process is sufficiently high to allow for a mapping of the momentum-resolved band structure. In contrast, indirect transitions, which involve the creation and annihilation of phonons, average out the momentum information of the electrons. The share of direct transitions is described by the Debye-Waller factor $DW$ which can be expressed by 
\begin{equation}
    DW (T) = \exp{(-\frac{1}{3}|\textbf{G}|^2\langle U^{2}(T)\rangle )},
\end{equation}
where $\textbf{G}$ is the reciprocal lattice vector---being large in HARPES experiments---that is involved in the transition and $\langle U^{2}(T)\rangle$ is the vibrational mean-squared atomic displacement at temperature $T$~\cite{gray2011}. For a single element compound, the latter is given by \mbox{$\langle U^{2}(T)\rangle = T \times (3\hbar^2/ M_a k_B \theta^2_D)$} with the atomic mass $M_a$, the Boltzmann constant $k_B$, and the Debye temperature $\theta_D$~\cite{shevchik1977}.
Since $\langle U^{2}(T)\rangle$ has to be small for a high yield of direct transitions, HARPES is thought to be applicable only to systems with stiff lattices and/or high atomic masses, and should be conducted at cryogenic temperatures. Indeed, it turns out that the high-$Z$ compound KTO is one of those systems which are---within the limits of the $k$ resolution set by the high kinetic energies---well-suited for HARPES, if cryogenic cooling is applied. Accordingly, the following measurements were conducted at a sample temperature of $T \approx$~55\,K. The results are shown in Fig.~\ref{fig:bandmap_kxky}.

Figure~\ref{fig:bandmap_kxky}(a) depicts a band map near the chemical potential, measured along the $\Gamma-X$-direction of KTO with $h\nu=3030$\,eV. Note that it is not corrected for the photon momentum $k_{\gamma}$. It covers electron emission angles within a range of $\Delta\theta=37^{\circ}$, measured in one shot with the wide angle lens of the spectrometer. Due to the high energies of the photoelectrons, one Brillouin zone~(BZ) corresponds to an angular range of only about $\Delta\theta=3^{\circ}$. Hence, several BZs along the $k_x$ direction are captured. The calculated zone edges of the KTO BZ scheme are marked by the dashed lines in the lower part of the figure. Because the high-energy incident x-rays are nearly parallel to the $k_x$ direction, there is a sizable transfer of momentum from the absorbed photons to the photoelectrons which results in an offset of the data with respect to the assigned zone scheme by a photoelectron momentum $k_{\gamma}$ of 1.53\,\AA$^{-1}$ along the $k_x$ axis. By coincidence, the corresponding shift is roughly the size of the $k_x$ width of one KTO BZ ($2\pi/a_\text{KTO}=1.57$\,\AA$^{-1}$).

From the band map it is evident that HARPES indeed allows us to observe the dispersion of the states near the chemical potential, which apparently follows the periodicity of the repeated zone scheme of KTO. Only at the BZ center spectral weight is observed, as expected from the vertex of the Ta~5$d$-derived conduction bands of KTO being located at the $\Gamma$-points~\cite{king2012,santander-syro2012}. The intensity of the different $\Gamma$ points seems to strongly depend on the specific zone probed as discussed further below. 

Figure~\ref{fig:bandmap_kxky}(b) shows an enlarged view of the band map around the $\Gamma$ point of the BZ~$(-1~0~18)$. The white dashed lines represent tight-binding calculations, taken from Figs.~2~(d) and 2~(f) of Ref.~\cite{santander-syro2012}. Density-functional theory (DFT) calculations for a model KTO slab  focusing on the effects of SOC on the conduction bands can be found in Appendix~\ref{app:SOCdim}. On the left-hand side of Fig.~\ref{fig:bandmap_kxky}(b), the tight-binding bands expected for $n$-doped bulk KTO are depicted. The right-hand side shows the first quantum well states~(QWSs) for the $d$ bands, hosting the surface electron system, under the assumption of a confining, wedge-shaped potential with an energetic depth of about $-0.5$\,eV and a spatial extension of a few unit cells in the $z$ direction. As can be seen, the limited $k$ resolution of the HARPES measurements does not allow us to resolve the specific details of the underlying $E(k)$ band dispersions. Nevertheless, a heavy \textit{bulk} $d_{yz}$ band, as depicted in the left half of Fig.~\ref{fig:bandmap_kxky}(b) to disperse well below the chemical potential towards the BZ boundary in the $k_x$ direction, is not seen in the experiment, suggesting that the $d_{yz}$ band actually is not bulk-like but disperses steeply to the chemical potential and beyond, similar to the case shown on the right side for the electron system strongly confined to the interface. Therefore, based on these observations, the q2DES appears to have a band structure similar to that of the surface q2DES on bare KTO and not to match that of $n$-doped bulk KTO.\\

Alternatively, the bulk $d_{yz}$ band could be present but not seen in the PES band map because its intensity is suppressed due to symmetry constraints for the photoemission matrix elements in the given measuring geometry~\cite{damascelli2003}. The photoemission intensity is proportional to
\begin{equation}
    |M_{fi}|^2 \propto |\langle \Psi_f | \textbf{\^{A}} \cdot \textbf{\^{p}} | \Psi_i \rangle |^2 \propto |\langle \Psi_f | \textbf{e} \cdot \textbf{\^{r}} | \Psi_i \rangle |^2,
\end{equation}
where $M_{fi}$ denotes the transition matrix element, $\textbf{e}$ is the unit vector along the polarization direction of the photon vector field $\textbf{\^{A}}$, $\textbf{\^{p}}$ and $\textbf{\^{r}}$ are the momentum and position operators, respectively, and $| \Psi_i \rangle$ ($| \Psi_f \rangle$) denote the initial (final) state. Note that the momentum operator can be expressed by the position operator using a commutation relation with the unperturbed Hamiltonian to obtain the commonly known form of the dipole operator $\textbf{e} \cdot \textbf{\^{r}}$~\cite{damascelli2003}.

An assessment of these effects in present experiments can be based on simple symmetry considerations that apply to our particular measuring geometry. The experimental setup for momentum mapping is sketched in Fig.~\ref{fig:bandmap_kxky}(d). The incident x-ray beam and the fan of photoelectrons entering the analyzer through the entrance slit lie in the mirror plane of the sample that is defined by the [001] and [100] directions. The photoelectron final states can be considered plane waves with even parity with respect to this mirror plane. Furthermore, the x-rays used in the experiments were linearly polarized within this mirror plane, i.e., the dipole operator also has even parity with respect to the mirror plane.

The orientation of the $d_{yz}$-states in question is depicted in the upper part of Fig.~\ref{fig:bandmap_kxky}(c). They have odd parities, thus the corresponding matrix element is zero ($M_{f,i} \propto \langle +|+|- \rangle = 0$, where $+$ and $-$ denote even and odd parity of the states and dipole operator, respectively), which indeed is pointing towards a substantial suppression of the photoemission intensity for the $d_{yz}$ band in Fig.~\ref{fig:bandmap_kxky}(b).

In contrast, as can be inferred from the lower part of Fig.~\ref{fig:bandmap_kxky}(c), the $d_{xz}$ orbitals in the same measuring geometry have even parity with respect to the mirror plane and thus should exhibit a non-vanishing photoemission yield ($M_{f,i} \propto \langle +|+|+ \rangle \neq 0$). The same is true if the sample is tilted about the $\phi$ axis to map the $k_y$ direction, since there are then no symmetry constraints on the dipole matrix element at all. Due to the four-fold rotational symmetry of the KTO(001) surface, the $d_{xz}$ band along the $k_y$ direction is electronically equivalent to the $d_{yz}$ band along the $k_x$ axis. Thus, if the bulk-like heavy $d_{yz}$ band initially discussed was not observed in the $k_x$ direction in our geometric setup, the symmetry-equivalent $d_{xz}$ band should be seen in our measurements, reaching far out along the $k_y$ direction.

Figure~\ref{fig:bandmap_kxky}(d) shows the $k_x-k_y$ map of the photoelectron intensity, integrated over the binding energy range from $-0.3$\,eV to 0.5\,eV, i.e., essentially the full Ta~5$d$ band width. The intensities thus represent the projected occupied states within the Fermi surface cuts rather than only the Fermi surface contours. It was assembled from measuring multiple band maps while varying the tilt angle $\phi$ [cf. schematic geometry on the right of Fig.~\ref{fig:bandmap_kxky}(d)]. The data are corrected for the photon momentum $k_{\gamma}$, and the BZs are labeled according to their Miller indices $h$ and $k$. The white grid indicates the calculated positions of the BZ-boundaries. The data confirms that the occupied part of the dispersive 5$d$ bands is located around the center of the BZs. The intensity distribution appears to be essentially isotropic within the experimental resolution, i.e., no intensity from a heavy $d_{xz}$ band, reaching out to the BZ boundary along the $k_y$ direction, is observed. Thus, the absence of a spectral signal of a heavy $d_{yz}$ band in Fig.~\ref{fig:bandmap_kxky}(b) is not ascribed to matrix element effects, but is likely due to the fact that the corresponding states indeed steeply disperse towards $E_F$, implying that the band structure of the buried electron system does not resemble the one of an $n$-doped, bulk-like KTO crystal. Instead, the electronic structure is assumed to be lifted by the confining potential in a manner similar to the KTO surface q2DES, as shown on the right-hand side of Fig.~\ref{fig:bandmap_kxky}(b). A similar observation is made with the $k_x-k_y$ map [Fig.~\ref{fig:kspace_microscope}(a)] and the $k_y$ band map [Fig.~\ref{fig:kspace_microscope}(b)(ii)] recorded with the momentum microscope at a higher kinetic energy.
\newpage
As mentioned in the discussion of the band map in Fig.~\ref{fig:bandmap_kxky}(a), also in Fig.~\ref{fig:bandmap_kxky}(d), apparent variations in the intensities and sizes of the occupied area at the centers of the displayed BZs are observed. While the intensity around the $\Gamma$ points of the BZs with $|h|<2$ is most pronounced, it appears entirely suppressed for those with $|h|=3$. These variations can be readily explained by the spherical cuts in $k$ space that are probed in (H)ARPES measurements at a fixed kinetic energy of the photoelectrons, as is illustrated  in Fig.~\ref{fig:bandmap_kxky}(e) for the cases of two- and three-dimensional free electron systems with cylindrical and spherical Fermi surfaces, respectively. For the latter case, the spherical cut does not intersect the Fermi surface equally in every BZ and thus low or no photoemission intensity is seen in some zones. Hence, this observation indicates a $k_z$ dependence of the electronic structure. An intensity dependence in the $k_z$ direction is often attributed to a three-dimensional character of the electron system. However, we would like to point out that already for confined electron systems with comparable sizes of Fermi wavelength and confinement length, the $k_z$-broadened quantum well states resemble the situation for an ideally three-dimensional electron system as schematically sketched in Fig.~\ref{fig:bandmap_kxky}(f), even more so if sizable experimental broadening is involved. This has been previously observed and reconciled with simulations of the intensity distribution using a Fourier analysis-based description of photoemission on confined electron systems~\cite{plumb2014,moser2018,strocov2018,santander-syro2020}.

\begin{figure*}
 \includegraphics[width=0.92\textwidth]{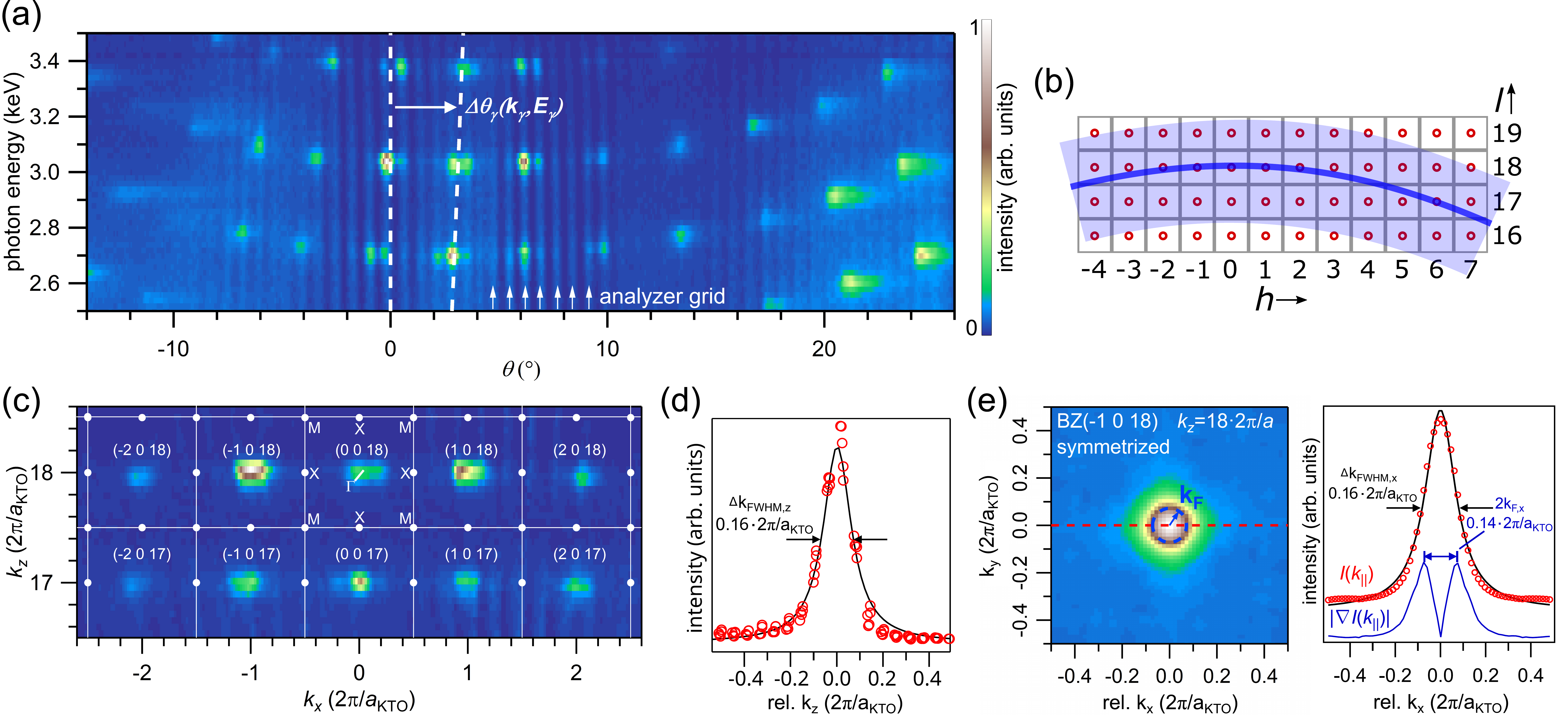}
 \caption{\label{fig:kxkz_evaluation}~(a)~Photoemission intensity near the Fermi energy of a dLAO/KTO sample with $t_{\text{dLAO}} = 1.6$\,nm at $k_y = 0$ as a function of photon energy and emission angle. (b) Reciprocal space covered by the HARPES measurements. (c)~Corresponding Fermi surface map of a 1.6\,nm dLAO/KTO sample in the $k_x-k_z$ plane. (d)~Intensity profile through the BZ center along the $k_z$ direction, averaged over five of the BZs in (c). (e)~Symmetrized $k_x-k_y$ map through the $\Gamma$ point of BZ~$(-1~0~18)$ and corresponding intensity profile through the BZ center (red). The modulus of the derivative of the profile (blue) yields an estimate for $\boldsymbol{k}_{F,x}$. The black lines in (d) and (e) are Lorentzian fits to determine the full width at half maximum (FWHM) of the intensity profiles.}
\end{figure*}
These intensity modulations across various BZs can also be seen in the PES map recorded over the photon energy range of 2500--3500\,eV in Fig.~\ref{fig:kxkz_evaluation}(a), i.e., with varying $k_z$. Here the step size of 20\,eV corresponds to $\approx0.1$\,\AA. The sample was aligned at a tilt angle $\phi = 0^\circ$, i.e., $k_y=0$.
The image is assembled from band maps over an angular range of about 40$^\circ$ along $k_x$ recorded at each photon energy. Like in the $k_x-k_y$ map in Fig.~\ref{fig:bandmap_kxky}(d), each pixel represents the integral intensity of a PES spectrum between $-0.3$ and 0.5\,eV. The corresponding $k$ space probed by this measurement is depicted in Fig.~\ref{fig:kxkz_evaluation}(b). It becomes clear that the intensity maxima occurring at the $\Gamma$ points again reflect the spherical sampling of the repeated zone scheme of an electron system that has a three-dimensional periodic pattern of occupied states in $k$-space of whatever origin.

In Fig.~\ref{fig:kxkz_evaluation}(a), the raw pattern is overlaid with vertical stripes of weaker intensity, which at first glance might be identified as features of the band structure that do not disperse along $k_z$. However, as indicated by the right dashed line, the $k_z$ axis in the image is tilted with respect to the photon energy axis due to the contribution of the energy-dependent photon momentum. In contrast, the stripes are running parallel to the photon energy axis. In fact, they are artifacts arising from the shadowing of a grid at the entrance of the analyzer lens.

To highlight the actual band structure of the sample, the raw data was normalized to the dispersionless background along the angular direction, before the emission angle $\theta$, and the photon energy was converted to $k_x$ and $k_z$, respectively. Figure~\ref{fig:kxkz_evaluation}(c) shows the central section of the map from the photon energy scans after the conversion. From the periodicity along $k_z$, we deduce an inner potential of $V_0=12\pm8$\,eV. The map confirms that for the perpendicular momentum component, the intensity is constrained to the centers of the BZs. The spectral weight nicely follows the repeated BZ scheme of bulk KTO. Thus, it is clear that the electronic structure of the buried electronic system of dLAO/KTO is apparently distinct from that of an  ideally two-dimensional system. In the following, we show that our $k$-resolved data is fully in line with the conclusions drawn above from depth profiling of the q2DES by angle-dependent HAXPES, namely, that the Ta~5$d$ electrons are indeed strongly confined to the interface.

To do so, we elaborate on the dimensional character of the interfacial charge carriers by analyzing the $k$ broadening of the PES data. The underlying idea is that, according to the Heisenberg principle, spatial confinement of a particle in the $z$ direction converts to an uncertainty in $k_z$, i.e., $k_z$ is no longer a good quantum number for the quantum state of an electron whose motion is confined in the $z$ direction. Instead, its quantum state in $k$-space is characterized by a probability amplitude $\Phi(k_z)$ which is the Fourier transform of the electron's wave function $\Psi(z)$. To estimate the dimensional character of the charge carriers along these lines, we compare in the following the Fermi surface broadening in $k_z$ to those in $k_x$ or $k_y$ directions \cite{strocov2018}. Before doing so, one should recall that there is also final state $k_z$ broadening at work which is inherent to the photoemission process. This is caused by the damping of the photoelectrons' wave functions in the solid with the broadening scaling as $1/\lambda_e$. Likewise, if the broadening is mainly induced by confinement, it is a measure for the depth extension of the confinement potential $L_z$. Now, an advantage of HAXPES is that the final state $k_z$ broadening is much smaller than for photoemission in the VUV or soft x-ray regimes due to the much higher penetration depth of the photoelectrons' wave function into the solid. For KTO the inelastic mean-free path is about 50\,{\AA} according to the TPP-2M formula \cite{tanuma1993,tanuma2004} and hence final state $k_z$ broadening is negligible compared to the experimental $k$ resolution of $\Delta k_\text{res} \approx 0.16$\,\AA$^{-1} \approx 0.1\times2\pi/a_\text{KTO}$. 
 
To determine first the $k_x$ broadening, the photoemission intensity map of BZ~$(-1~0)$ at $k_z = 18\times2\pi/a_\text{KTO}$ [see Fig.~\ref{fig:bandmap_kxky}(d)] was symmetrized according to the four-fold rotational symmetry of the KTO substrate to eliminate the influence of geometrical matrix element effects. The result is displayed in Fig.~\ref{fig:kxkz_evaluation}(e). Since we do not resolve the Fermi surface contours, we take the entire extension of the Fermi surface as the benchmark, defined by the full width at half maximum (FWHM) of the cross-section profile along the red line. A comparison with that along $k_{z}$ [see Fig.~\ref{fig:kxkz_evaluation}(d)]---note that here the intensity profile was obtained by averaging over five of the BZs in Fig.~\ref{fig:kxkz_evaluation}(c)---reveals that they are essentially equal, viz., $\Delta k_{\text{FWHM},x} \approx \Delta k_{\text{FWHM},z} = 0.16 \times 2\pi/a_{\text{KTO}}$. Thus, no confinement induced broadening of the perpendicular wave-vector is deducible from the measurements. Inversely, for a broadening to be deducible from the measurements, it had to be larger than the experimental $k$ resolution. Hence, since no additional broadening along $k_z$ can be resolved, it must be smaller than the experimental resolution, which sets a lower limit of $1/\Delta k_\text{res}\approx 10$\,{\AA} on the extension of the wave functions of the electrons, forming the q2DES, along the $z$ direction, fully consistent with the estimate from the previous HAXPES depth profiling for the q2DES extension of 20\,{\AA}.

Therefore taking this value as granted, there is another independent way to estimate the dimensional character of the interface electron system using a simple argument. The dimensionality of an electron system is not solely determined by its extension in the confinement direction, $L_z$, but rather by the ratio of the extension and the Fermi wavelength of the electrons, $\lambda_{F,z}$, with the dimensional character becoming three-dimensional if $L_z \gg \lambda_{F,z}$ \cite{rogers1984,rogers1987,dassarma2011}. Here $L_{z} \approx 20$\,{\AA} and $\lambda_{F,z} = \frac{2 \pi}{k_{F,z}} \approx 57$\,{\AA}, which gives a ratio of $\approx 0.35$, placing the electron system in favor of a 2D character. For the determination of $k_{F,z}$, we set---based on the findings above---$k_{F,z}=k_{F,x}$, and take the modulus of the gradient of the photoemission intensity map as measure for $k_{F,x}$ [see Fig.~\ref{fig:kxkz_evaluation}(e)]. This method was shown in Ref.~\cite{straub1997} to give accurate results for the Fermi vector even if large energy intervals are used for integration to generate a Fermi surface map. 

From our analysis, we arrive at the following picture: Under confinement along the $z$ direction, the three-dimensional band structure transforms into discrete two-dimensional QWSs, whose wave-vector component $k_z$ is not well-defined anymore, inducing a broadening of the intensity distribution
in this direction. However, if the depth extension of the electron system is still large enough in comparison to $\lambda_{F,z}$, the intensity measured in HARPES along $k_z$ can still be limited to the part of momentum space, where the corresponding (hypothetical) three-dimensional Fermi surface would be located. The value of $\approx 0.35$ for the ratio of confinement length and Fermi wavelength appears to place the electron system in dLAO/KTO in this \textit{quasi}-two-dimensional regime between the pure 2D and 3D limits.

\section{Conclusion}

In conclusion, we have established the successful preparation of a highly spin-orbit coupled, metallic q2DES at the (001)-oriented surface of KTaO$_3$ by PLD of a dLAO film at low oxygen pressures. Compared to the prototypical SrTiO$_3$-based electron systems that are also induced by oxygen depletion, the Hall mobilities, reaching up to 7000\,cm$^2$V$^{-1}$s$^{-1}$ at cryogenic temperatures and up to 30\,cm$^2$V$^{-1}$s$^{-1}$ at room temperature, are remarkably high, which we ascribe to the different effective electron masses and depth extensions of the conduction channels.

It has been demonstrated that angle-resolved hard x-ray photoemission spectroscopy is suitable to study the momentum-resolved electronic structure of this electron system, only a few nanometers thin and buried under a protecting layer. The occupied states around the $\Gamma$ points in momentum space follow the repeated zone scheme of three-dimensional KTaO$_3$. Nevertheless, detailed band maps along high-symmetry directions lack characteristic bulk bands but are in line with band calculations for a q2DES at the KTaO$_3$ surface that develops QWSs. These findings are in line with the results of core-level depth profiling and allow us to estimate the extension of the interfacial electron system from the experimental broadening of the electrons' wavevector component perpendicular to the surface. Our combined results demonstrate that the spatial confinement of the electronic interface system is only marginally smaller than its Fermi wavelength, thereby placing its dimensional character in the \textit{quasi}-two-dimensional regime between the pure 2D and 3D limits.
 
\begin{acknowledgments}
G. Sangiovanni would like to thank Domenico Di Sante for advice and useful comments.\\
This paper was supported by the Deutsche Forschungsgemeinschaft (DFG, German Research Foundation) through the W\"urzburg-Dresden Cluster of Excellence on Complexity and Topology in Quantum Matter ct.qmat (EXC 2147, Project ID No. 390858490), the Collaborative Research Center SFB 1170 ToCoTronics (Project ID No. 258499086) in W\"urzburg and the Transregional Collaborative Research Center TRR 173 Spin+X in Mainz (Project ID No. 268565370). We thank the Bundesministerium f\"ur Bildung und Forschung (BMBF) through the project EffSpin-HAXPES (Project ID No. 05K16WWA and No. 05K16UMC) for additional funding. Diamond Light Source (Didcot, UK) is gratefully acknowledged for beamtime at beamline I09 under Proposals No. SI-15200, No. SI-14432, and No. SI-11394 as well as the Deutsches Elektronen-Synchrotron DESY (Hamburg, Germany) for beamtime at beamline P22 under Proposals No. I-20181092 and No. I-20181063.
\end{acknowledgments}

M. Z. and M. Schmitt contributed equally to this work.

\appendix

\section{Ta~4$f$ core-level spectra and irradiation-induced doping}
\label{sec:corelevel}

\begin{figure}
 \includegraphics[width=0.45\textwidth]{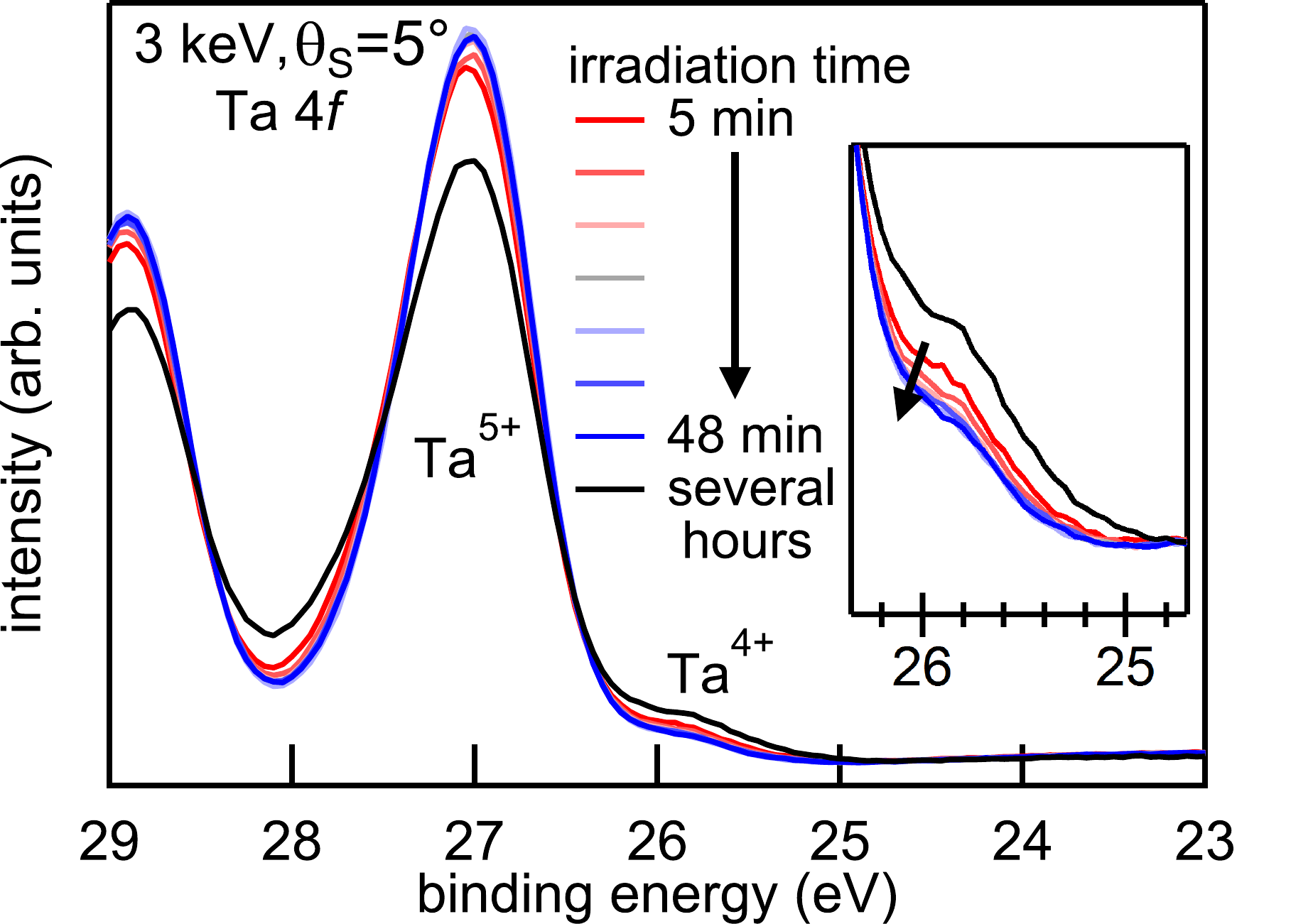}
 \caption{\label{fig:irradiation_depthprof}~Evolution of the Ta~4$f$ core level under irradiation with hard x-rays ($h\nu$ = 3\,keV) in grazing incidence geometry at $\theta_s$ = 5$^\circ$.}
\end{figure}

Third generation storage rings like the Diamond Light Source or PETRA III feature synchrotron radiation with extraordinary high brilliance, which can induce changes of the sample properties during a measurement. Various experiments have shown that perovskite oxides are prone to modifications of their chemical and electronic structure upon irradiation with highly intense photon beams, in particular through the creation of oxygen vacancies \cite{meevasana2011,dudy2016,gabel2017} that effectively act as electron donors. Likewise, bare KTO tends to release oxygen when irradiated with photons in the ultraviolet range, as has been demonstrated in synchrotron-based ARPES experiments \cite{king2012}. Therefore, we monitored the stability of the dLAO/KTO electron system by repeating measurements of the Ta~4$f$ core level during continuous irradiation with hard x-rays. Core-level spectra of cations can be used as a probe of additional electrons in the valence shell through the concomitant energetic shift of photoemission spectral weight to lower binding energies as the oxidation state is reduced and the core-hole potential is better screened.

Figure~\ref{fig:irradiation_depthprof} depicts Ta~4$f$ core-level spectra recorded repeatedly over a period of about 45\,min on a previously unirradiated sample spot. The measurement was conducted in a geometry similar to that for the HARPES measurements. The spectra represent the collected intensity integrated over an emission angle range of $\theta=-12^{\circ}$ to $19^{\circ}$ with respect to the surface normal of the sample. The spectra are normalized to total Ta~4$f$ intensity. It is directly seen from the inset that the Ta$^{4+}$ contribution at lower binding energies, signaling the q2DES, is initially decreasing with irradiation time, which is likely caused by photoinduced reactions of contaminants and the removal of surface adsorbates \cite{scheiderer2015}. The ratio of the Ta$^{4+}$ signal to the total Ta~4$f$ intensity decreases from 6.7\,\% to a minimum of around 5.2\,\%. The black line is an additional spectrum taken after continuous measurements over several hours at the same sample spot. Apparently, the $n$ doping into the Ta~5$d$ shell meanwhile recovered and increased to even higher values than observed initially.  

\section{Influence of spin-orbit coupling on the dimensionality of the electron system}\label{app:SOCdim}

\begin{figure}
	\includegraphics[width=0.45\textwidth]{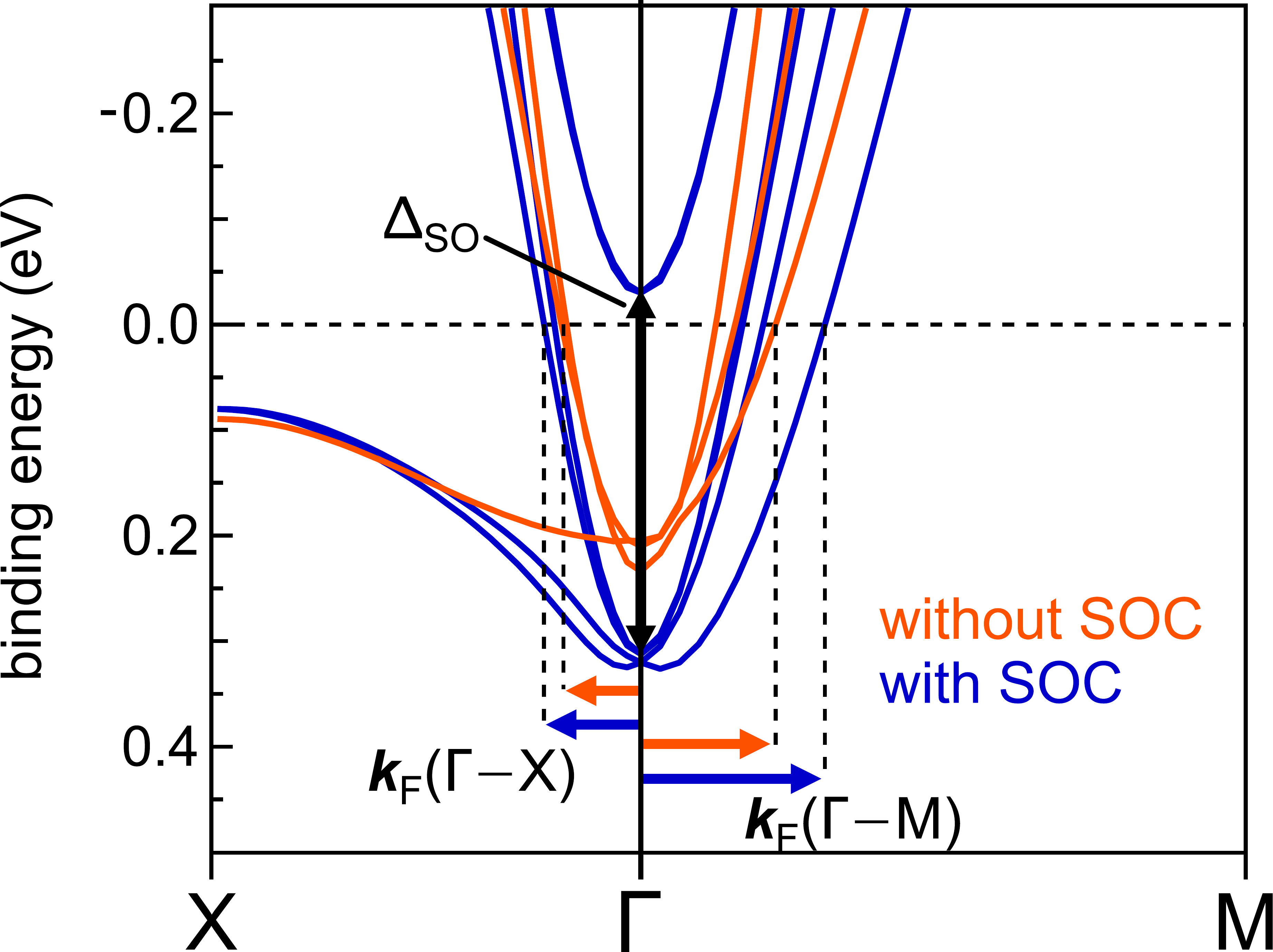}
	\caption{\label{fig:LDA}~DFT-based band structure calculations for a $1\times1\times10$ slab of KTO with a TaO$_2$-terminated surface without and with SOC taken into account. Shown is a comparison of the Fermi wave-vectors $\boldsymbol{k}_F$ along the high-symmetry directions of the surface BZ of both cases. $\boldsymbol{k}_F$ is enlarged when SOC is taken into account.}
\end{figure}

The dimensionality of an electron system is determined by the ratio of its extension and the Fermi wavelength of the electrons~\cite{rogers1984,rogers1987,dassarma2011}. Taking into account interactions beyond non-relativistic band theory may lead to a renormalization of the electronic structure and thus can have a distinct influence on this ratio. An example, especially for KTO, are SOC-induced changes of the conduction bands.

To illustrate the potential impact of these effects in the case of KTO, we compare the bandstructure of a TaO$_2$-terminated 1x1x10 KTO slab along the [001] direction with and without SOC. The DFT calculations have been done using the generalized gradient approximation as parametrized in the PBE-GGA functional \cite{kresse1996,kresse1999,perdew1996}, without the explicit inclusion of oxygen vacancies and without structural relaxation. We have checked that, allowing for lattice relaxations of the first two unit cells of the slab yields qualitatively equivalent results. Figure~\ref{fig:LDA} illustrates this comparison, focusing on the surface Ta~5$d$-derived band structure close to the Fermi level. The most obvious change is the SOC-induced splitting of otherwise degenerate bands. For a more detailed analysis, we inspect the three lowest Ta~5$d$-derived bands for the case without SOC (orange) and with SOC (blue). When SOC is included, one of these three bands is shifted upward above the chemical potential, while the two other bands are lowered, preserving charge neutrality. Thus, SOC modifies KTO from a three to an effective two conduction-band system. Additionally, at the KTO surface, SOC induces a Rashba-like splitting that lifts the spin degeneracy of each of the bands. The respective Fermi wave-vectors $\boldsymbol{k}_F$ are also indicated in the two plotted high-symmetry directions of the surface BZ. Due to the shift of two bands to lower energies, $\boldsymbol{k}_F$ is increased by about 20\,\% and 35\,\% in the $\Gamma-X$- and $\Gamma-M$-directions, respectively.

To be more illustrative, a larger $\boldsymbol{k}_F$, i.e., a smaller $\lambda_F$, effectively results in more QWSs being occupied in the system than without SOC, which increases the three-dimensional character of the electronic states. However, the relative size of the changes in $\boldsymbol{k}_F$ of a few tens of percent suggests that the dimensional character of the conduction electrons in KTO is not fundamentally altered compared to similar electron systems with smaller SOC.\\

\section{Momentum-resolved hard x-ray time-of-flight microscopy}\label{app:ToF-mic}

The concept of a momentum microscope is to simultaneously image the three-dimensional ($E_{kin}$,$k_x$,$k_y$) distribution of photoelectrons using ToF energy recording in combination with an objective lens, resulting in much higher count rates and a larger field of view in momentum space than when using a hemispherical analyzer. Here, a newly developed momentum-resolved microscope was used, which is equipped with a multi-mode front lens that can operate in the classical extractor mode and also in a field-free mode (advantageous for three-dimensional structured samples) as well as in a space-charge suppression mode (important for intense radiation, e.g., from free-electron lasers) \cite{medjanik2019}. For the experiments of this study, it was set up at the high brilliance hard x-ray beamline P22 of the synchrotron PETRA III (DESY, Germany) \cite{schlueter2019}. Using the 40-bunch mode of PETRA III (granting a bunch separation of 192\,ns) and a delay-line detector with a diameter of 80\,mm, the covered $k$ space can be up to 25\,\AA$^{-1}$. In this experiment, the energy resolution of 150\,meV is set by the photon band width. The best achievable energy and momentum resolutions of the microscope are 19\,meV and 0.025\,\AA$^{-1} \approx 0.02\times2\pi/a_\text{KTO}$, respectively \cite{medjanik2019}.

\begin{figure*}
 \includegraphics[width=0.9\textwidth]{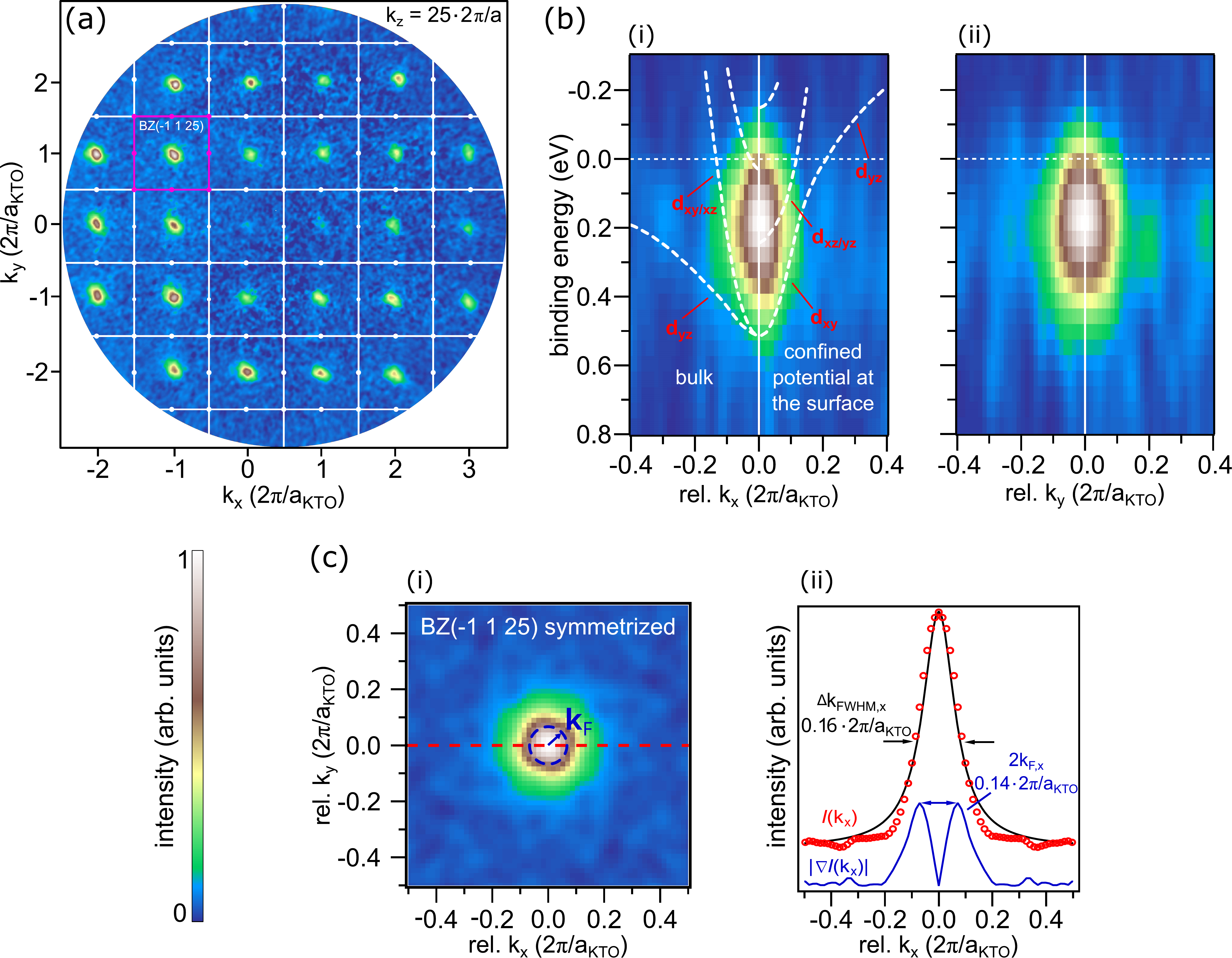}
 \caption{\label{fig:kspace_microscope}~(a)~Fermi surface map in the $k_x-k_y$ plane of a dLAO/KTO sample with $t_{\text{dLAO}}=1.6$\,nm measured at $h\nu$ = 5177\,eV ($k_z=25\times2\pi/a_{\text{KTO}}$) with an angle-acceptance of about $\pm$7.2$^\circ$ in $k_x$ and $k_y$ directions, together with a schematic drawing of the corresponding $k$ space. A square-like arrangement of the intensities around the $\Gamma$ points is clearly visible. The data is not corrected for the photon momentum. (b)~Band maps through the $\Gamma$ point of BZ~$(-1~1~25)$ along (i) the $k_x$ and (ii) the $k_y$ directions. Both band maps do not show a bulk-like $d_{yz}$ or $d_{xz}$ band, respectively. White dashed lines are tight-binding calculations from Ref.~\cite{santander-syro2012} [see also Fig.~\ref{fig:bandmap_kxky}(b)]. (c)(i)~Symmetrized $k_x-k_y$ map of the region around the $\Gamma$ point of BZ~$(-1~1~25)$ and (ii) corresponding line profile through the BZ center (red). The blue line represents the modulus of the derivative of the red profile. The black line is a Lorentzian fit to determine the width of the intensity profile. The data is in very good agreement with those taken at lower photon energies with the hemispherical analyzer (see main text).}
\end{figure*}

A Fermi surface map in the $k_x-k_y$ plane of a dLAO/KTO sample with $t_{\text{dLAO}}=1.6$\,nm recorded with this microscope is shown in Fig.~\ref{fig:kspace_microscope}(a), together with a schematic drawing of the corresponding $k$ space. The map was measured at a photon energy of $h\nu = 5177$\,eV, corresponding to a value for $k_z$ of $25\times2\pi/a_{\text{KTO}}$. The field of view, determined by an acceptance angle of $\pm7.2^\circ$, covers up to 26~BZs, with the intensity around the $\Gamma$ points showing a square-like arrangement reflecting the cubic KTO lattice. The acquired data represent a spherical cut through the periodic zone scheme at a final-state energy defined by photon energy, binding energy and inner potential, as discussed in Ref. \cite{medjanik2017}, and therefore, not all $\Gamma$ points appear with the same high intensity (cf. Sec.~\ref{sec:HARPES}).

In Fig.~\ref{fig:kspace_microscope}(b), two band maps---representing cuts through the $\Gamma$ point of BZ~$(-1~1~25)$---are shown for the (i) the $k_x$ and (ii) the $k_y$ direction. Neither a heavy $d_{yz}$ band is visible along the $k_x$ direction, nor a heavy $d_{xz}$ band in the $k_y$ direction, in agreement with the claim of a surface-confined q2DES made above.

Figure~\ref{fig:kspace_microscope}(c)(i) shows the symmetrized intensity map around the $\Gamma$ point of BZ~$(-1~1~25)$ with (ii) the corresponding line profile $I(k_{x})$ (red) and the modulus of its derivative $|\nabla I$(k$_{x}$)$|$ (blue) in very good agreement with the data acquired with the hemispherical analyzer.

%\bibliographystyle{apsrev4-2}
%\bibliography{Library}

%

\end{document}